\documentclass[pre,aps,twocolumn,showpacs,floatfix,superscriptaddress]{revtex4-2}
\usepackage{graphicx}
\usepackage{bm}
\usepackage{color}
\usepackage{amssymb,amsmath}
\usepackage{ulem}
\usepackage{ragged2e}
\usepackage{adjustbox}
\usepackage{multirow}
\usepackage{booktabs}
\usepackage[table]{xcolor}
\usepackage{tcolorbox}
\usepackage{tabularx}
\usepackage{array}
\usepackage{placeins}
\usepackage{colortbl}
\usepackage{hyperref}
\hypersetup{colorlinks=true, citecolor=blue, urlcolor=blue, linkcolor=blue}
\tcbuselibrary{skins}
\definecolor{Salmon}{RGB}{250,128,114}

\newcolumntype{Y}{>{\raggedleft\arraybackslash}X}

\tcbset{tab1/.style={\centering, fonttitle=\bfseries\small,fontupper=\normalsize\sffamily,
		colback=yellow!10!white,colframe=red!75!black,colbacktitle=Salmon!40!white,
		coltitle=black,center title,freelance,frame code={
			\foreach \n in {north east,north west,south east,south west}
			{\path [fill=red!75!black] (interior.\n) circle (3mm); };},}}

\tcbset{tab2/.style={enhanced,fonttitle=\bfseries,fontupper=\small\sffamily,
		colback=yellow!10!white,colframe=red!50!black,colbacktitle=Salmon!40!white,
		coltitle=black,center title}}
\begin{document}	
\title{Propagation of extreme events in multiplex neuronal networks}
\author{R. Shashangan}
\email{shashang.physics@gmail.com}
\affiliation{Department of Nonlinear Dynamics, Bharathidasan University, Tiruchirappalli 620024, Tamil Nadu, India}
\author{S. Sudharsan}
\email{hari.sudharsan32@gmail.com}
\affiliation{Physics and Applied Mathematics Unit, Indian Statistical Institute, Kolkata 700108, India}
\author{Dibakar Ghosh}
\email{diba.ghosh@gmail.com}
\affiliation{Physics and Applied Mathematics Unit, Indian Statistical Institute, Kolkata 700108, India}
\author{M. Senthilvelan}
\email{velan@cnld.bdu.ac.in (Corresponding Author)}
\affiliation{Department of Nonlinear Dynamics, Bharathidasan University, Tiruchirappalli 620024, Tamil Nadu, India}
\email{velan@cnld.bdu.ac.in}
\begin{abstract}
In previous studies, the propagation of extreme events across nodes in monolayer networks has been extensively studied. In this work, we extend this investigation to explore the propagation of extreme events between two distinct layers in a multiplex network. We consider a two-layer network, where one layer is globally coupled and exhibits extreme events, while the second layer remains uncoupled. The interlayer connections between the layers are either unidirectional or bidirectional. We find that unidirectional coupling between the layers can induce extreme events in the uncoupled layer, whereas bidirectional coupling tends to mitigate extreme events in the globally coupled layer. To characterize extreme and non-extreme states, we use probability plots to identify distinct regions in the parameter space. Additionally, we study the robustness of extreme events emergence by examining various network topologies in the uncoupled layer. The mechanism behind the occurrence of extreme events is explored, with a particular focus on the transition from asynchronous states to a fully synchronized excitable state. For numerical simulations, we use nonidentical FitzHugh-Nagumo neurons at each node, which captures the dynamical behavior of both coupled and uncoupled layers. Our findings suggest that extreme events in the uncoupled layer emerge through the gradual disappearance of disorder, accompanied by occasional bursts of synchronized activity. Results obtained in this work will serve a starting point in understanding the dynamics behind the propagation of extreme events in real-world networks.
\end{abstract}
\maketitle
\section{Introduction}
\par Extreme event is a generalized term used to describe a class of events that are rare and recurrent with destructive aftermath. Tsunami, earthquakes, epileptic seizures, epidemics, share market crashes, and power blackouts are some examples of extreme events (EE) in nature and human-made systems \cite{Kantz}. Whenever these events occur, we can observe the magnitude of that respective event to appear as an outlier having deviated far from its median. Indeed, because of this peculiar property, they are christened as EE. Epilepsy is one such outlier event occurring as a result of abrupt, abnormal and excessive synchronization among the neurons inside the brain. Despite synchronization being mandatory for various essential cognitive functions, neural plasticity and memory processes \cite{Fell}, when it happens abruptly and excessively, it becomes lethal. Although several technological improvements have made possible the diagnosis and treatment of this disorder, the fundamental reason for such a mystifying behavior is still unknown. One possible solution to get deeper insights about the underlying mechanism, responsible factors, propagation, and spread, is by studying them in coupled dynamical systems. 

\par By and large, in coupled dynamical systems, at the point of such abnormal synchronization, the time series of the collective observable of the system (for example the mean field of the state variables) is found to qualify peak over threshold (POT) criterion \cite{Walkers}. Such classified EE and the corresponding return interval should fit with any one of the non-Gaussian distributions \cite{Extremes}. In coupled dynamical systems, the presence of interaction between the oscillators is found to induce EE due to the occurrence of one or more of the following phenomena, namely (i) imperfect phase synchronization \cite{Ansmann}, (ii) excitability of proto events \cite{Ansmann}, (iii) instability of out of phase synchronization \cite{Mishra}, (iv) attractor bubbling \cite{Oria}, (v) bubble transition and blowout bifurcation \cite{Feudel}, (viii) intermittency through saddle-node bifurcation \cite{Vaibhav}, and (ix) intermittent cluster synchronization \cite{sudharsan1}. These have become the responsible precursors for the EE in networks with models like, FitzHugh-Nagumo (FHN) \cite{Ansmann}, Hindmarsh-Rose model \cite{sudharsan1} Li\'enard type system \cite{Kapitaniak}, Josephson junction \cite{Kapitaniak}, Stuart-Landau oscillators \cite{AP}, chaotic maps \cite{SNC}, moving agents of chaotic oscillators \cite{SNC1} and pendulum \cite{tapas}. 
\par One important property of the EE is its nature to propagate and spread, especially when the network has layered architectures. This spread may disrupt the functioning of the entire network. Previously, the propagation of the EE from one node to another has been studied depending on the coupling strength and the amount of parameter heterogeneity \cite{Kapitaniak}. In addition, the propagation of the EE from the high-degree node to the low-degree node has been studied, depending on the coupling strength \cite{ray2022}. Both studies \cite{Kapitaniak,ray2022} were carried out in a monolayer network, and the propagation of EE from one node to another was studied, which allowed us to understand the spread of these events in monolayer networks. Owing to the presence of layered architectures in several real-world applications, it is equally important to study the propagation of EE from one network to another. To our best search of literature, we found that Chen et al. \cite{Seager} have previously studied EE in multi-layer interdependent social networks using the statistical approach. Several works in the theory of coupled oscillators have proven the role of interlayer interaction in significantly improving the rhythmicity of neuronal networks \cite{Srilena}, stimulating synchronization \cite{Eckehard}, controlling chimera states \cite{Mikhay,Rybalova}, promoting solitary states \cite{Solitary}, inducing stochastic and coherence resonance \cite{Nadezhda,Anna}. However, no study in the theory of coupled oscillators, so far has dealt neither with the study of the propagation of EE from one network to another nor has unearthed the relation between the emergence of EE and the interlayer interaction. Upon extending our studies on this aspect we uncovered an important role played by the interlayer coupling on the emergence and mitigation of EE in a two-layer multiplex network. In particular, we investigate the following three questions: (i) How a multilayer neuronal network facilitates the propagation of EE in a unidirectional setting? (ii) How does the information in one layer influence the EE in other layer in a bidirectional setting?, and (iii) what is the role of different intralayer coupling topologies in the propagation and mitigation of extreme events?

\par To answer these questions we consider a two-layer multiplex network where each node follows the dynamics of the FHN neuron model and analyze the role of interlayer coupling strength on the emergence of EE in this network. One layer is considered to contain nodes that are globally coupled exhibiting EE while the other layer is considered uncoupled. With this network configuration, we specifically study how EE emerge in an uncoupled layer due to interlayer coupling and thereafter extend to local, nonlocal and global architectures. Specifically, we study the effect of interlayer coupling in two specific cases depending on the direction of the interaction: (i) Unidirectional - when the interaction is only from the coupled to the uncoupled layer, and (ii) Bidirectional - when the interaction is directed both the ways. For this, we consider two types of interlayer coupling functions, namely (i) electrical and (ii) chemical synapses. 

\par In the literature, the coupling strength of a single layer network has been proved to be pivotal in the generation and mitigation of EE in both lower dimensional and higher dimensional coupled systems \cite{Walkers}. In particular, in the case of lower dimensional systems, bidirectional coupling \cite{Ansmann}, time-delayed coupling \cite{Feudel}, excitatory coupling, and inhibitory coupling \cite{Mishra} are found to induce EE as a result of coupling. Whereas, in Ref. \cite{sudharsan}, interplay between coupling strength and frequency mismatch is found to induce EE in a two-coupled chaotic oscillator. While on contrary, threshold activated coupling \cite{Ikeda} suppresses EE in a two-coupled maps. Extending to the case of higher dimensional monolayer networks, EE are reported to emerge through parametric coupling \cite{Promit}, local coupling \cite{ARS,sudharsan1}, time varying coupling \cite{SLK} and static and dynamic random link coupling \cite{Random}. Further, Roy and Sinha \cite{ARS} have shown  that global coupling plays the dual role in both emergence and mitigation of EE. In this direction, the present work is first of its kind in the avenue of EE in multilayer networks. While not only extending the analysis to multilayer networks, through this work, we also report an important finding that unidirectional interlayer coupling induces EE in an uncoupled layer. While the bidirectional interlayer coupling suppresses the already existing EE in the coupled layer. We also found that this phenomenon is robust to other coupling topologies like local, nonlocal and global implemented in the place of uncoupled layer. No study to the best of our knowledge has reported the role of interlayer coupling on the propagation of EE from one network to another. In addition, we report for the first time the dual role played by the interlayer coupling strength in the propagation and mitigation of EE based on the direction of information transfer. Our work will serve as a foundation and forerunner for exploring and comprehensively understanding the emergence and mitigation of EE in multiplex networks.

We present our results in the following way. In Sec.~\ref{model}, we describe the model and the network topology considered for our analysis and also define the measures used for our analysis. In Sec.~\ref{unidir}, we detail the dynamics and the propagation of EE when the interlayer coupling is unidirectional. Whereas, in Sec.~\ref{bidir}, we discuss the dynamics and the mitigation of EE when the interlayer coupling is bidirectional. Further, in Sec.~\ref{robust_stat}, for electrical coupling, we present the robustness of the results under different network topologies and also statistically characterize the observed EE using statistical measures.  Next, we discuss the mechanism for the propagation and mitigation in Sec.~\ref{mechanism}. Penultimately, in Sec.~\ref{conclusion}, we discuss and conclude the work. Finally, we present the statistical measures and mechanism of chemical coupling case in the Appendix. 

\section{Model and Measures} 
\label{model}
\par The two-layer network topology that we consider in this work is portrayed in the schematic Fig. 1. It is a multiplex network where each layer has same number of neurons and the neurons in layer-1 are connected only to its replica in layer-2. The neurons inside layer-1 are uncoupled whereas the neurons within layer-2 are globally coupled. As our prime motive is to investigate the propagation of EE, we choose layer-2 to  be exhibiting EE and then investigate how it propagates to the uncoupled layer-1. For this, we use FHN neuron model in each node in both the layers. The mathematical form for the considered two-layer network with FHN neuron is \\
\begin{subequations}
\textbf{Layer-1:}
\begin{eqnarray}
\dot{x}_{i,1}  &=& x_{i,1}(a - x_{i,1})(x_{i,1} - 1) - y_{i,1} +  \eta_{1}h_{1}(x_{i,1},x_{i,2}),    \nonumber \\
\dot{y}_{i,1}  &=& b_{i}x_{i,1} - cy_{i,1},  \label{ly1} 
\end{eqnarray}
\textbf{Layer-2:}
\begin{eqnarray}
\dot{x}_{i,2}  &=& x_{i,2}(a - x_{i,2})(x_{i,2} - 1) - y_{i,2} \nonumber \\&& + \epsilon_{2}\sum_{j=1}^{N} B_{ij}(x_{j,1} - x_{i,1}) + \eta_{2}h_{2}(x_{i,1},x_{i,2}), \nonumber \\
\dot{y}_{i,2}  &=& b_{i}x_{i,2} - cy_{i,2},  \quad i = 1, 2,...,N. \label{ly2}
\end{eqnarray}	
\end{subequations}

\begin{figure}[!h]
	\centering
\includegraphics[width=1.0\linewidth]{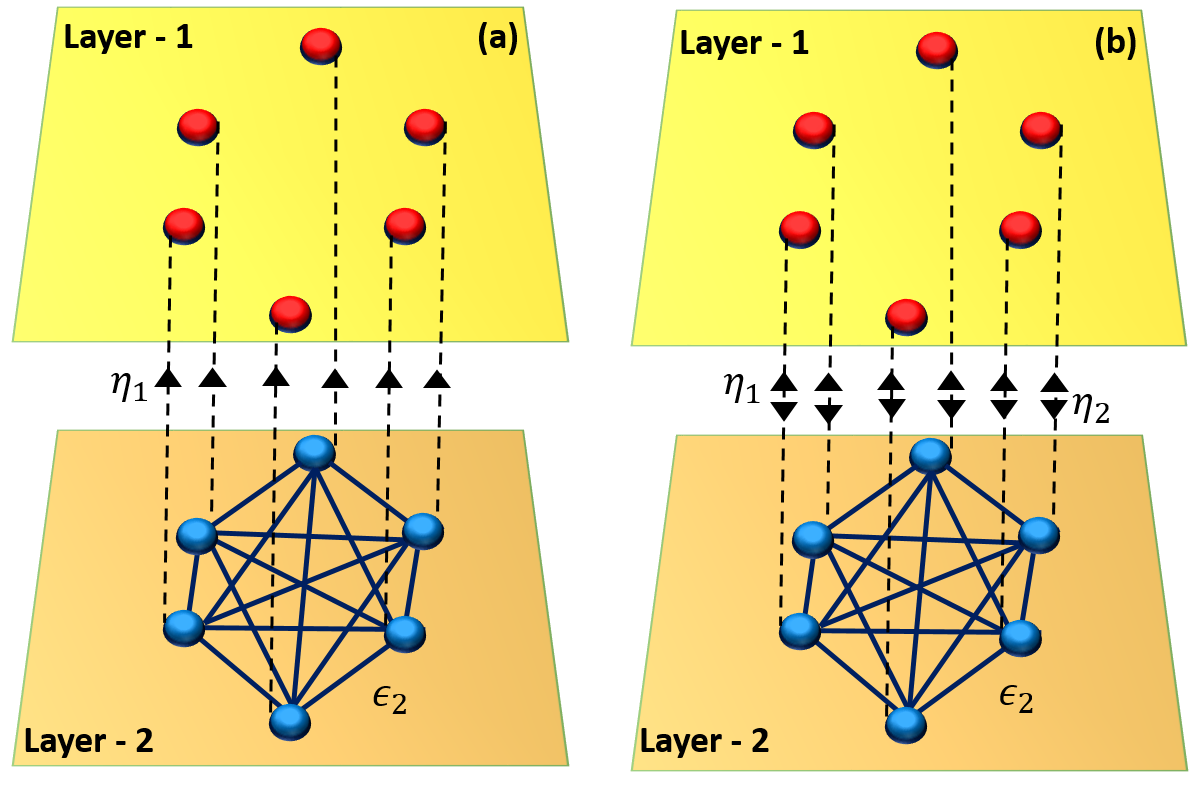}
	\caption{Schematic diagram of the two-layer FHN network with (a) unidirectional and (b) bidirectional interactions. Neurons in layer-1 (red filled circles) are uncoupled whereas neurons in layer-2 (blue filled circles) are globally coupled.}
\end{figure}
\begin{figure}[!h]
\centering
\includegraphics[width=0.5\textwidth]{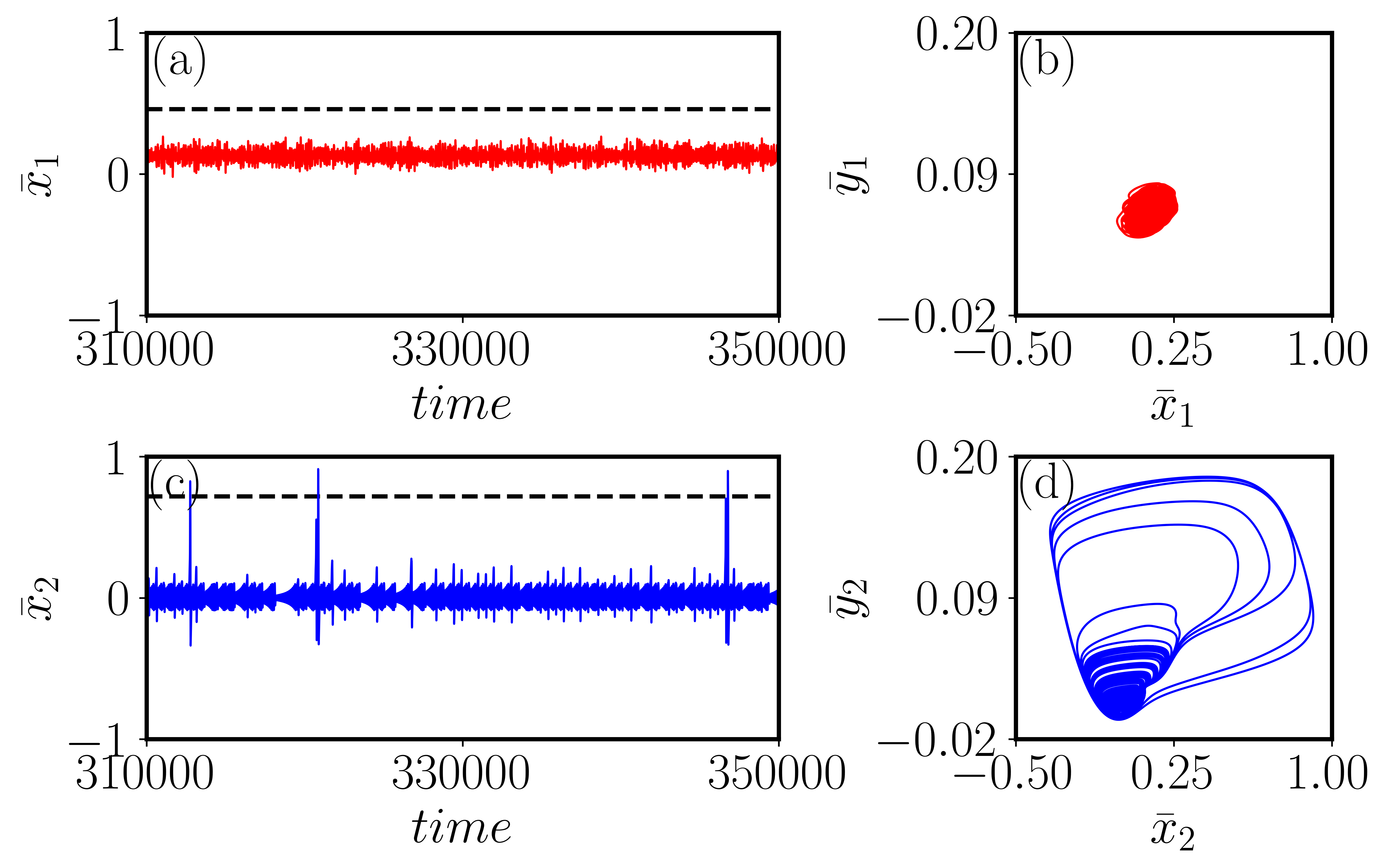}
\caption{Plots (a) and (b) are the time series and phase portrait of the collective observable ($\bar{x}$) in layer-1 while plots (c) and (d) are the time series and phase portrait of layer-2. Layer-1 exhibits disordered dynamics while layer-2 exhibits EE at $\eta_1=\eta_2=0.0$. The black dashed line corresponds to the threshold of EE.}
\end{figure}  
\par Here $x$ and $y$ represents the membrane potential and recovery variable of a neuron, respectively. The parameter $a$ describes the shape of the cubic polynomial, while $b$ and $c$ describe the coupling strength of feedback from the membrane potential and relative timescale between membrane potential and recovery variable respectively. $\epsilon_{2}$ is the intralayer coupling strength of layer-2. $B_{ij}$ is the adjacency matrix of the intralayer coupling of layer-2 which takes the value $B_{ij} = 1$ if the nodes $i$ and $j$ are connected\textbf{,} otherwise $B_{ij} = 0$.  In our case, we consider globally coupled intralayer coupling in the layer-2, so $B_{ij} = 1$ for $i\neq j$. Further, $\eta_{1}$ and $\eta_{2}$ are the interlayer coupling strengths from layer 2 to 1 and from layer 1 to 2, respectively. In order to make the layer-2 exhibit EE, as in Ref. \cite{Ansmann}, we fix the values of the parameters as $a = -0.02651$, $c = 0.02$ while the parameter $b_{i}$ is distributed from the expression $b_{i} = 0.006 + 0.008(\frac{i - 1}{N - 1})$. We fix $N=101$ in each layer. Throughout the work, the network topology, parameters' value, and the dynamics of layer-2 are fixed. The parameters' value of oscillators in layer-1 are the same as that of them in layer-2. Oscillators in layer-2 are globally coupled with each other and exhibit extreme events whereas in layer-1, the oscillators are totally uncoupled.  Later in Sec. \ref{robust_stat}, we take up the other coupling schemes in layer-1. 
\par The dynamics of the collective observable $\bar{x}_{l}=\displaystyle \frac{1}{N} \sum_{j=1}^{N}x_{j,l},~\text{and}~\bar{y}_{l}=\displaystyle\frac{1}{N}\sum_{j=1}^{N}y_{j,l} \quad (l=1, 2)$ in layers 1 and 2 separately are given in Figs. 2(a-d). This dynamics corresponds to $\eta_{1}=\eta_{2}=0$ where there is no transfer of information between the layers. Layer-2 exhibits EE while the dynamics of layer-1 is chaotic. We can observe from Fig. 2(c) that the time series comprises of numerous small and occasional large amplitude oscillations and these large oscillations can be viewed as the long excursion of the trajectory from the bounded chaos in Fig. 2(d). To identify EE from other events, we use peak-over-threshold (POT) method. By this method, whenever the time evolution of the system exceeds a certain threshold value then the corresponding observable is said to exhibit an extreme behavior. This definition for the threshold is intuitively apparent from the fact that the EE fall far away from the central tendency making it to appear in the tail of the probability distribution.
\par Mathematically, the threshold \cite{Walkers} is calculated using 
	\begin{equation}
		x_{th} = \langle \bar{x}_{i} \rangle + n\sigma_x, \quad n \in \mathbb{R}~\backslash\{0\}~\text{and}~n>1,
	\end{equation}
where $\langle \bar{x}_{i}\rangle$ is the mean of the peaks of the collective observable ($\bar{x}_{i}$), $\sigma_x=\sqrt{\big(\bar{x}_{i})^2  - \langle\bar{x}_{i}\rangle^{2}}$ is the standard deviation. Here, $n$ is a positive real number that determines the rareness of EE. The rarity of the EE is highly influenced by this number. Throughout this work, we fix $n=8$ and estimate $x_{th}$, as this choice of $n=8$ denotes that the emerged EE are more rarer. Since EE from the neuronal context is characterized by abnormal and excessive synchronization of neurons in the network, we analyze the nature of the intralayer ($S^{intra}_{E}$) and interlayer synchronization ($S^{inter}_{E}$) errors \cite{Bera} using
	\begin{equation}
		S^{intra}_{E} = \displaystyle \lim_{T\to \infty} \frac{1}{T} \int_0^ T \sum_{j=2}^{N}\frac{\lvert \lvert x_{j,1}(t) - x_{1,1}(t) \rvert\rvert}{N-1} dt,
        \label{sync_1}
    \end{equation}
    \begin{equation}
		S^{inter}_{E} = \displaystyle \lim_{T\to \infty} \frac{1}{T} \int_0^ T \sum_{j=1}^{N}\frac{\lvert \lvert y_{j,2}(t) - x_{i,1}(t) \rvert\rvert}{N} dt. 
        \label{sync_2}
	\end{equation}  
\par In this considered network topology, two factors are crucial for the transfer of information from one layer to another, (i) interlayer and (ii) intralayer coupling strength. Since the $\epsilon_{2}=0.00128$ is fixed, we first study the effect of $\eta_{1,2}$ and later the effect of $\epsilon_1$. For $\eta_{1,2}$ there are two possibilities, (i) unidirectional and (ii) bidirectional. The unidirectional form of transmission denotes the passage of information only in the forward direction (A$\rightarrow$B or B$\rightarrow$A), whereas the bidirectional form denotes the passage of information in both forward and backward direction (A$\leftrightarrows$B).  

\section{Unidirectional Interaction ($\eta_1\neq 0,\eta_2=0$)}
\label{unidir}
\par First, we discuss the unidirectional interaction where exchange of information is only in one direction from layer-2 to layer-1. There will be no transfer of information from layer-1 to layer-2. We first analyze the layers interacting with electrical coupling and then later with chemical coupling. 
\subsection{Electrical coupling}
\par In this case, we set $\eta_{1} \neq 0$, $\eta_{2} = 0$, and $\epsilon_{2} = 0.00128$ in Eq. (1a). The unidirectional electrical interaction in Eq. (1) is given by
	\begin{equation}
		h_{1}(x_{i,1},x_{i,2}) = x_{i,2} - x_{i,1}.
	\end{equation}
	\begin{figure*}[!ht]
		\centering
		\includegraphics[width=1.0\textwidth]{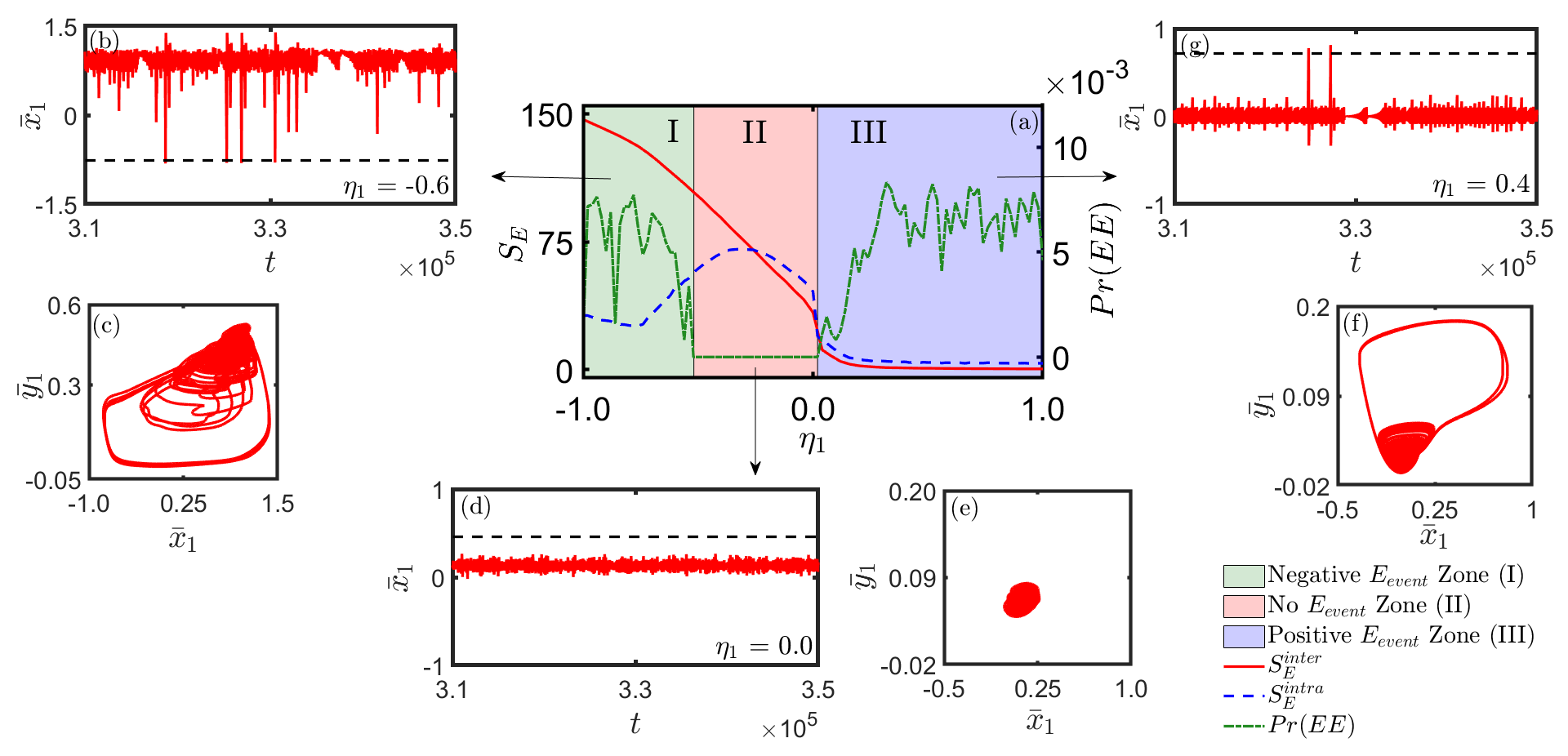}	
		\caption{(a) Variation of synchronization error and probability plot of system (1) for the uncoupled layer-1 in unidirectional electrical coupling by varying $\eta_{1}$. The interlayer synchronization error ($S^{inter}_{E}$), intralayer synchronization error ($S^{intra}_{E}$) and probability of occurrence of EE  ($Pr(EE)$) are indicated by red solid, blue and green dashed lines respectively. Plots (b,c at $\eta_{1}=-0.6$), (d,e at $\eta_{1}=0.0$) and (g,f at $\eta_{1}=0.4$) corresponds to time series and phase portraits of layer-1 dynamics in zones I, II and III. The black dashed line corresponds to the threshold.}
		\label{se_ts_1}
	\end{figure*}
\par We start to investigate the role of $\eta_{1}$ in the emergence of EE in layer-1 by visualizing its dynamics. For this purpose, we discern the zones of $\eta_{1}$ and observe the EE dynamics, synchronization error and the probability of occurrence of EE in Fig.~\ref{se_ts_1}. The probability of EE occurrence is estimated by normalizing the count of EE with the total number of peaks used in the calculation of $x_{th}$. Specifically, in Fig.~\ref{se_ts_1}(a), the red solid, the blue dashed and the green dashed lines indicate interlayer synchronization error ($S^{inter}_{E}$), intralayer synchronization error ($S^{intra}_{E}$) and the probability of occurrence of EE ($Pr(EE)$) respectively of layer-1. Further, Figs.~\ref{se_ts_1}(b,d,g) represent the time series and correspondingly Figs.~\ref{se_ts_1}(c,e,f) represent the phase portraits respectively for $\eta_{1} = -0.6$, $0.0$ and $0.4$. Upon varying $\eta_{1}$, it is found that EE emerge in layer-1 after a certain value of $\eta_{1}$. This is true for both excitatory and inhibitory $\eta_{1}$. Examining the $\eta_{1}$ parameter space for EE, three different zones are evidenced as shown in Fig.~\ref{se_ts_1}(a). Zones I and III are the zones of EE while Zone-II is a non-extreme event zone. This is confirmed by zero probability for the emergence of EE in the layer-1 in Zone-II (peach) and nonzero probability in Zone-I (green) and Zone-III (violet) zones. A major difference can be observed in the EE emerging in the excitatory ($\eta_{1} > 0$) and inhibitory ($\eta_{1} < 0$) zones. In the excitatory zone, extreme maximum occurs whereas in the inhibitory zone, extreme minimum occurs. In zone-I ($\eta_{1} \in (-1.0,-0.518)$), EE occur in layer-1 in the negative spatial direction (negative EE) when there is sufficiently higher inhibitory coupling. The large amplitude valleys crossing the threshold in Fig.~\ref{se_ts_1}(b) corresponds to negative EE which can be observed as the long excursion of the trajectory from the bounded chaotic region in Fig.~\ref{se_ts_1}(c). Next in zone II ($\eta_{1} \in (-0.518,0.02)$) no EE occur. It is confirmed from the Figs.~\ref{se_ts_1}(d,e), where there are neither large amplitude valleys/peaks crossing the threshold in Fig.~\ref{se_ts_1}(d) nor long excursion of the trajectory from the bounded chaotic region in Fig.~\ref{se_ts_1}(e). Further in the Zone-III ($\eta_{1} = 0.02$ to $1.0$) for very small value of $\eta_{1}$, EE emerge in the positive spatial direction which is opposite to the occurrence of EE in the green zone. This can be visualized from the Figs.~\ref{se_ts_1}(f,g). Here the layer-1 mimics the dynamics exactly as in layer-2. We observe that while tuning $\eta_{1}$ from $-1.0$ to $1.0$, the spatial direction of emergence of EE changes from negative to positive. 
\par We calculate the intralayer and interlayer synchronization errors using Eqs. (\ref{sync_1}) and (\ref{sync_2}) and analyze the synchronization error within and between the two layers. From the Fig.~\ref{se_ts_1}(a), we can observe that the $S^{inter}_{E}$ (red solid curve) is maximum for $\eta_{1} = -1.0$ and decreases gradually while increasing the $\eta_{1}$ leading to zero $S^{inter}_{E}$. That is while varying $\eta_{1}$, the layers transit into the synchronization regime from the desynchronized regime and finally attain complete synchronization. On the other hand, $S^{inter}_{E}$ (blue dotted curve) is non-zero at $\eta_{1} = -1.0$ starts to slightly decrease up to $\eta_{1} \approx -0.75$. Then there is an increase and decrease in the error up to $\eta_{1} \approx 0.02$ and after that it saturates to a nonzero value. This shows that neurons inside layer-1 transits into synchronization regime but does not attains complete synchronization within themselves. When the coupling is inhibitory, EE emerge where the two layers desyncrhonized (non-zero $S^{inter}_{E}$ in Fig.~\ref{se_ts_1}(a)) whereas during excitatory coupling, EE emerge amidst synchronization between the layers (zero $S^{inter}_{E}$ in Fig.~\ref{se_ts_1}(a)). The same scenario occurs in case of intralayer synchronization ($S^{intra}_{E}$) as well. In addition when the layers are in synchronized, we get the exact dynamics in layer-1 as in layer-2 (EE occurs in positive spatial direction). The complete mechanism of the emergence of EE will be detailed in Sec. \ref{mechanism}.   
\subsection{Chemical coupling}
\par Now we consider the interlayer coupling through chemical synapse. The interlayer chemical coupling is a nonlinear coupling \cite{Majhi} in the form
	\begin{eqnarray}
		h_{1}(x_{i,1},x_{i,2}) = (V_{s} - x_{i,1})\Biggl(\frac{1.0}{1.0+e^{(-\lambda(x_{i,2} - \Theta_{s}))}}\Biggl).
	\end{eqnarray}
\par
\begin{figure*}[!ht]
		\centering
		\includegraphics[width=1.0\textwidth]{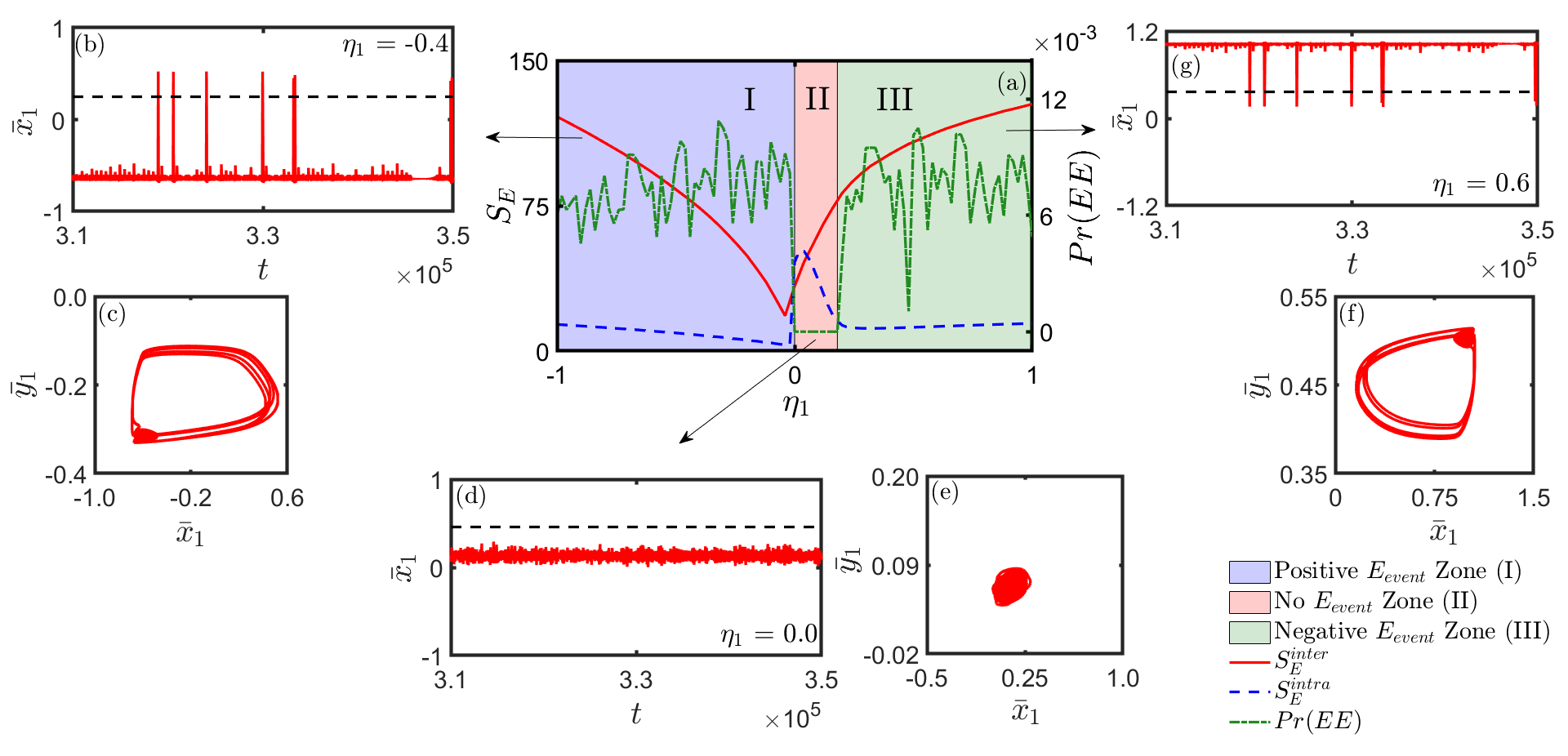}	
		\caption{Variation of synchronization errors $S_E$ and probability of occurrence $Pr(EE)$ by changing the interlayer coupling strength of system (1) for the uncoupled case in unidirection chemical coupling is displayed in plot (a). The time series and phase portraits of layer-1 dynamics in zones I, II and III corresponds to the plots ($b,~c$ at $\eta_{1}=-0.4$), ($d,~e$ at $\eta_{1}=0.0$) and ($g,~f$ at $\eta_{1}=0.6$). The black dashed horizontal line corresponds to the threshold. The legends and color schemes are same as in Fig.~\ref{se_ts_1}.}
		\label{se_ts_2}
	\end{figure*}

 Here, $V_{s} = 2.0$, $\Theta_s=-0.25$, $\lambda=10$. We use Eq. (6) in Eq. (1) and analyze in the same fashion as in the previous subsection, the nature of emergence of EE in the layer-1. In this regard, Fig.~\ref{se_ts_2} illustrates the dynamics of EE, synchronization errors, probability of emergence of EE in layer-1 for chemical coupling. In particular, Figs.~\ref{se_ts_2}(b,d,g) represent the time series and Figs.~\ref{se_ts_2}(c,e,f) represent the phase portraits respectively for $\eta_{1} = -0.4$, $0.0$ and $0.6$, respectively. Color schemes and legends are the same as in Fig.~\ref{se_ts_1}. Upon inferring the dynamics of EE in layer-1, a major difference is observed between electrical and chemical coupling cases. The zones of the positive and negative EE are completely interchanged. That is the positive EE occur in the inhibitory zone and negative EE occur in the excitatory zone. The large amplitude peaks crossing the threshold in dynamics of Zone I ($\eta_{1} \in (-1.0,-0.02)$) in Fig.~\ref{se_ts_2}(b) corresponds to positive EE and is visualized as the long excursion of the trajectory from the bounded chaotic region in Fig.~\ref{se_ts_2}(c). In Zone II  ($\eta_{1} \in (-0.02,0.2)$) no EE occurs and can be confirmed from the Figs.~\ref{se_ts_2}(d,e) where there are neither large amplitude peaks crossing the threshold (in Fig.~\ref{se_ts_2}(d)) nor long excursion of the trajectory from the bounded chaotic region (in Fig.~\ref{se_ts_2}(e)). In Zone-III ($\eta_{1} \in (0.2,1.0)$), the EE occur due to excitatory coupling in the spatial direction opposite to the EE in Zone-I (Figs.~\ref{se_ts_2}(f,g)). We observe that while varying $\eta_{1}$ from $-1.0$ to $1.0$, there is a shift in the spatial direction of the emergence of EE from positive to negative. In electrical coupling, positive EE emerge in layer-1 exactly identical to the layer-2. But in the chemical coupling, the EE are not identical between layer-1 and layer-2. Further, the range of zero $Pr(EE)$ ($\eta_{1} \in (-0.02,0.2)$, Zone-II (red)) for the emergence of EE in layer-1 is less when compared with the unidirectional electrical case (Fig.~\ref{se_ts_1}(a)).
\par In Fig.~\ref{se_ts_2}(a), the red solid line indicates $S^{inter}_{E}$ which is maximum for $\eta_{1} = -1.0$ and it starts to decrease as increasing the interlayer coupling strength up to $-0.04$. Then again from $\eta_{1} \approx -0.03$, the error begins to increase and reaches a maximum value. In this unidirectional chemical interaction, the two layers are desynchronized and the synchronization between them is not attained at all. Further, $S^{intra}_{E}$ of layer-1 (blue dashed line in Fig.~\ref{se_ts_2}(a)) starts to decrease up to $\eta_{1} \approx -0.02$ and then there is increase in the error and reaches a maximum value at $\eta_{1} = 0.02$. Then the error starts to decrease and saturates to a very small nonzero value. This indicates that the neurons inside layer-1 are not synchronized with each other. On the contrary with electrical coupling, when the coupling is chemical, EE emerges in both excitatory and inhibitory regimes where the two layers are in desynchronized (non-zero $S^{inter}_{E}$ and $S^{intra}_{E}$ in Fig.~\ref{se_ts_2}(a)) throughout the parameter-$\eta$ space.

\section{Bidirectional Interaction $(\eta_1=\eta_2\neq 0)$}
\label{bidir}
\par Until this point, we have investigated the role of interlayer coupling when the interaction is unidirectional from layer-2 to layer-1. Now, we analyze the emergence of EE when the interaction is bidirectional, that is mutual sharing of information takes place between the layers. For bidirectional coupling we set $\eta_{1}$ = $\eta_{2}=\eta \neq 0$ and analyze system (1) in the same way as taken up in Section.~\ref{unidir}, first electrical and then chemical coupling.
\subsection{Electrical Coupling}
The bidirectional electrical interactions are given by
	\begin{eqnarray}
		h_{1}(x_{i,1},x_{i,2}) = x_{i,2} - x_{i,1}, \nonumber \\
		h_{2}(x_{i,1},x_{i,2}) = x_{i,1} - x_{i,2}.
	\end{eqnarray}
\par Here $h_{1}(x)$ represents information transfer from layer-2 to layer-1 and $h_{2}(x)$ denotes the transfer of information from layer-1 to layer-2. When we begin to examine the dynamics of layer-1, we surprisingly found that, in the bidirectional case, EE does not emerge in layer-1. In contrast, EE already existing in layer-2 is mitigated completely. So, in the present case, we analyze the collective dynamics of layer-2 instead of layer-1. On increasing the interlayer coupling strength, owing to the bidirectional nature of the coupling, the signal incoming from layer-1 affects the dynamics of the oscillators in layer-2 thereby mitigating the extreme events already existing in layer-2.
	\begin{figure*}[!]
		\centering
		\includegraphics[width=1.0\textwidth]{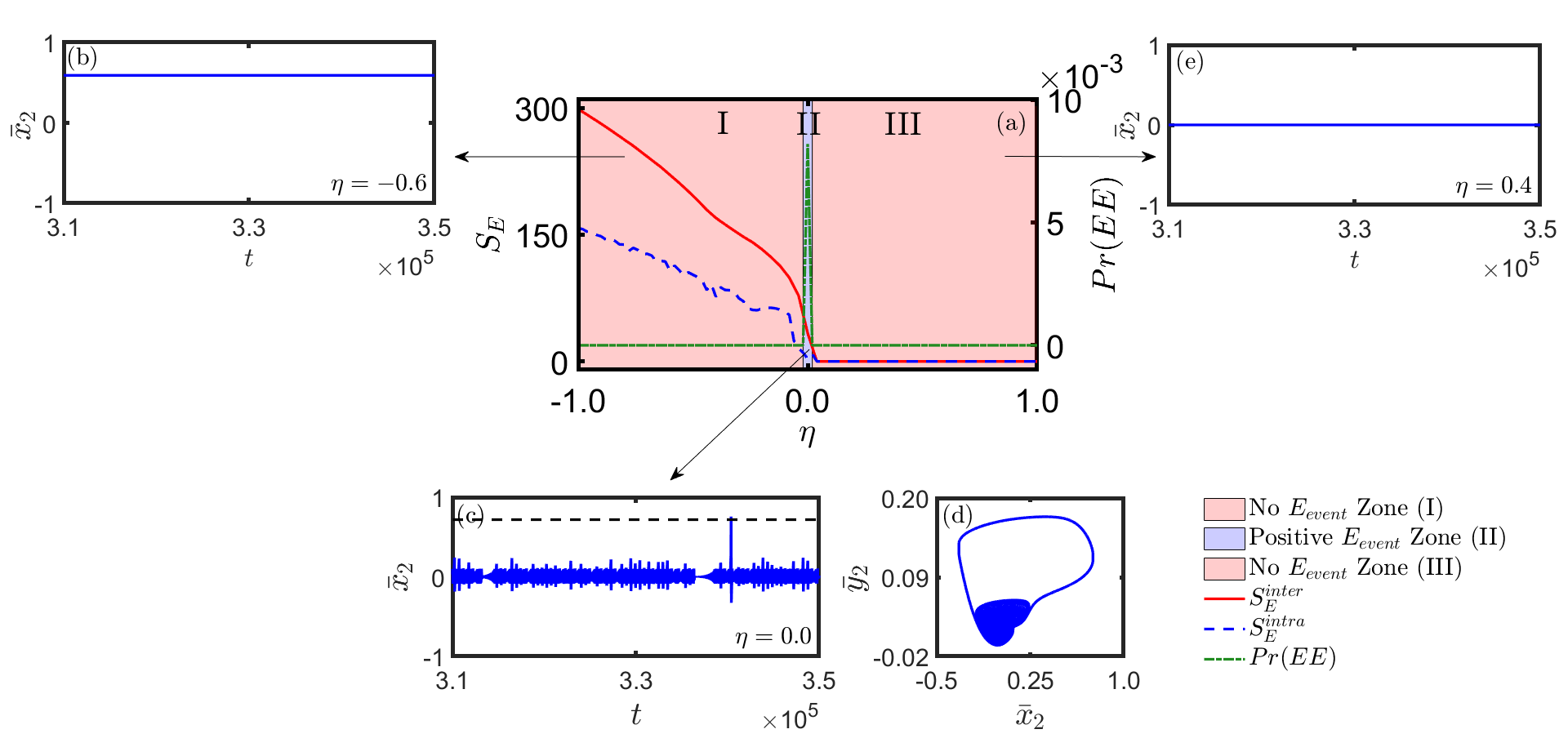}
		\caption{Variation of $S_{E}$ and $Pr(EE)$ against $\eta$ of layer-2 in system (1) is illustrated in (a) for the uncoupled case in bidirectional electrical coupling. The time series and phase portraits of layer-2 in zone II corresponds to the plots ((c-d) at $\eta=0.0$). The black dashed line corresponds to the threshold of EE. $S^{inter}_{E}$, $S^{intra}_{E}$ and $Pr(EE)$is identified by red solid line, blue and green dashed lines, respectively.}
		~\label{se_ts_3}
	\end{figure*}
\par The plots in Fig.~\ref{se_ts_3} describes the dynamics, synchronization errors and the probability of EE in layer-2. Figures~\ref{se_ts_3}(c) and ~\ref{se_ts_3}(d) depict the presence of EE in the time series and phase portraits respectively for $\eta = 0.0$. While Figs.~\ref{se_ts_3}(b) and \ref{se_ts_3}(e) portray the absence of EE in time series for $\eta = -0.4$ and $\eta = 0.6$. In both excitatory and inhibitory cases, after mitigation the dynamics transits into a steady state. At microlevel, the steady state for the inhibitory case is of two-cluster type where oscillators split into two clusters differing in amplitude whereas in case of excitatory coupling, the steady state is of single cluster where all the oscillators undergoes an amplitude death (AD). In Fig.~\ref{se_ts_3}(a), the red solid line, blue and green dashed lines indicate $S^{inter}_{E}$, $S^{intra}_{E}$ and $Pr(EE)$ of layer-2, respectively. When the interaction is bidirectional, the EE in the source layer (layer-2) is mitigated. Further from Fig.~\ref{se_ts_3}(a), we can observe that EE exist only for a short range which is from $\eta\in(-0.02$,$0.02)$ (blue Zone-II) and in the rest of the zones (red Zones-I and III) EE disappear.  
\par While varying the synaptic strength from inhibitory to excitatory ($\eta = -1.0$ to $1.0$), $S^{inter}_{E}$ decreases and becomes zero at $\eta\approx0.05$ and continues to be the same till $\eta=1.0$. In the same manner, the $S^{intra}_{E}$ decreases and saturates to zero. All these results can be observed from Fig.~\ref{se_ts_3}(a). In this case, both the layers evolve into a completely synchronized state and the neurons in layer-2 are also completely synchronized. Further, for inhibitory coupling, the mitigation of EE occur when the two layers desynchronized (non-zero $S^{inter}_{E}$ and $S^{intra}_{E}$ in Fig.~\ref{se_ts_3}(a)) while for excitatory coupling, the mitigation occurs when the layers are synchronized (zero $S^{inter}_{E}$ and $S^{intra}_{E}$ in Fig.~\ref{se_ts_3}(a)). 

\subsection{Chemical coupling}
\par Next, we consider bidirectional chemical interaction similar to the unidirectional case, with the only difference that, in the present bidirectional case, the network will have both the chemical functions $h_1,h_2$ as given by,
	\begin{eqnarray}
		h_{1}(x_{i,1},x_{i,2}) = (V_{s} - x_{i,1})\Biggl(\frac{1.0}{1.0+e^{(-\lambda(x_{i,2} - \Theta_{s}))}}\Biggl), \nonumber\\
		h_{2}(x_{i,1},x_{i,2}) = (V_{s} - x_{i,2})\Biggl(\frac{1.0}{1.0+e^{(-\lambda(x_{i,1} - \Theta_{s}))}}\Biggl).
	\end{eqnarray}
	\begin{figure*}[!ht]
		\centering
		\includegraphics[width=1.0\textwidth]{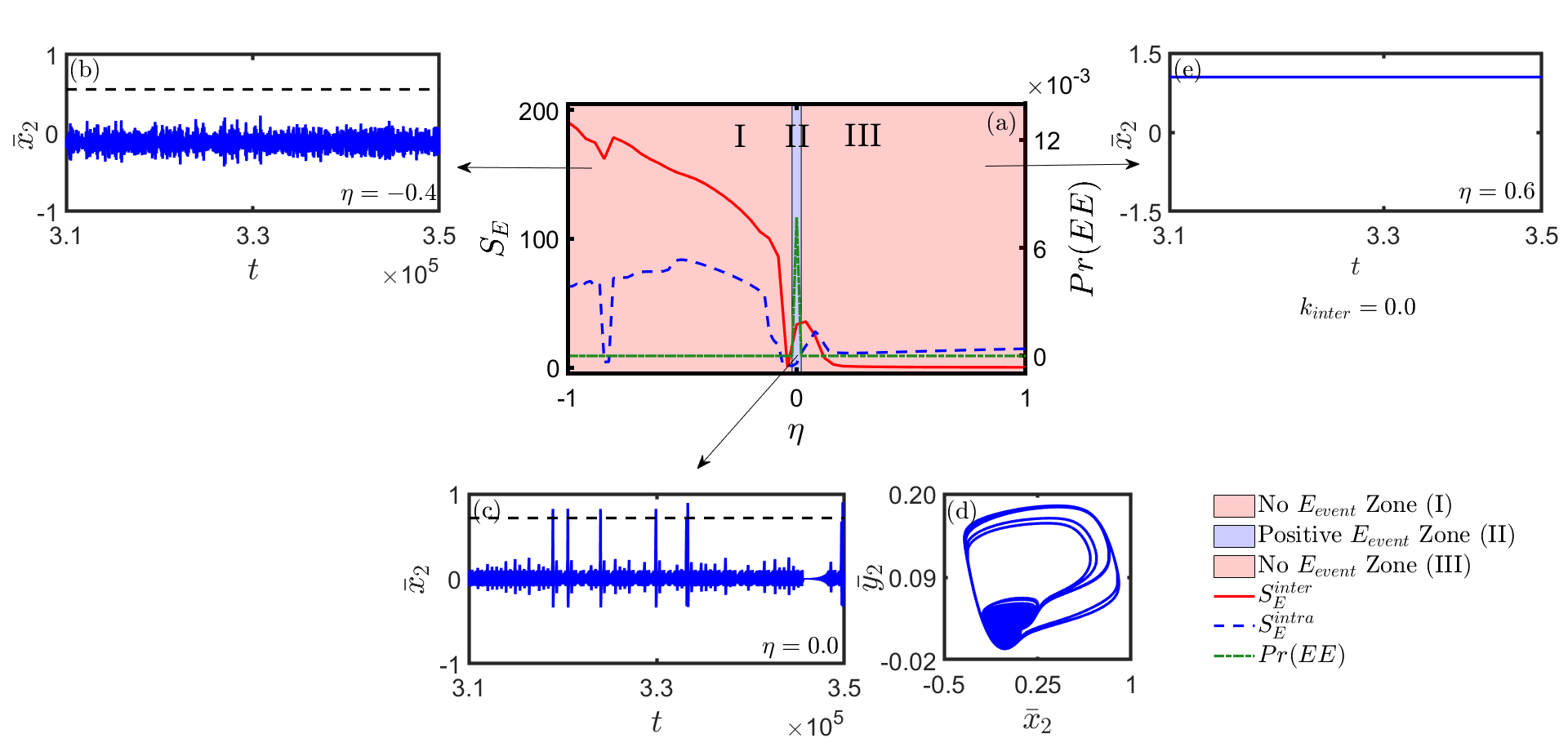}
		\caption{The change in $S_{E}$ and $Pr(EE)$ over $\eta$ of system (1) for the uncoupled case in bidirectional chemical coupling is shown in plot (a). Plots ((c-d) at $\eta=0.0$) corresponds to time series and phase portraits of layer-2 dynamics in zone II. The black dashed line corresponds to the threshold. Legends and color schemes are same as in Fig.~\ref{se_ts_3}.}
		~\label{se_ts_4}
	\end{figure*}
\par In this case, the EE dynamics, synchronization and the probability of EE in layer-2 can be observed as depicted in Fig.~\ref{se_ts_4}. Figures ~\ref{se_ts_4}(c) and \ref{se_ts_4}(d) represent the time series and the phase portraits respectively for $\eta = 0.0$. Due to bidirectional interaction, the EE in the source layer (layer-2) is mitigated as in the electrical case and the probability of occurrence of EE in layer-2 (shown by green dashed line in Fig.~\ref{se_ts_4}(a)) is also similar to the electrical case. A major difference is that only chaotic state prevails and no steady state occur for inhibitory coupling. The steady state for the excitatory case is of a single cluster (same amplitude for all oscillators). The $S^{inter}_{E}$ (red solid line in Fig.~\ref{se_ts_4}(a)) starts to decrease and saturates to zero till $\eta=1.0$. The two layers transit into synchronized regime from desynchronized regime. The $S^{intra}_{E}$ (blue dashed line) decreases and saturates to a very small nonzero value. The neurons in layer-2 are completely desynchronized. On the whole, when the coupling is bidirectional and chemical, the synchronization inside the layers and between the layers occur in the same way as in the electrical bidirectional case.

\section{Robustness in propagation and mitigation and statistics of extreme events}
\label{robust_stat}
\par So far, the intralayer network topology of layer-1 was considered to be uncoupled. To check whether the change in network topology of layer-1 has any effect on the propagation of EE or not, we vary the network topology of layer-1 and then carry out the analysis of extreme events under both unidirectional and bidirectional interlayer coupling. The equation for layer-1 with intralayer coupling topology now reads
\begin{eqnarray}
\centering
\dot{x}_{i,1}  &=& x_{i,1}(a - x_{i,1})(x_{i,1} - 1) - y_{i,1} \nonumber \\&&+ \frac{\epsilon_{1}}{2p}\sum_{j=1}^{N} A_{ij}(x_{j,1} - x_{i,1}) + \eta_{1}h_{1}(x_{i,2},x_{i,1}),    \nonumber \\
\dot{y}_{i,1}  &=& b_{i}x_{i,1} - cy_{i,1}.
\end{eqnarray}
\par Here $A_{ij}$ is the intralayer adjacency matrix of layer-1 which is defined as $A_{ij}=1~~\text{for} ~~ 0 < |i-j|\le p$ and $A_{ij}=0$ otherwise, where $p$ is the coupling radius that determines the nature of topology in the layer-1 intralayer coupling. For $p=1$, the interaction is of local type, when $p=\frac{N-1}{2}$, the interaction is of global type and when the value of $p$ lies in the range $1<p<\frac{N-1}{2}$, the interaction is of nonlocal type where $N$ is the number of nodes (odd) in the layer. When $\epsilon_{1}=0$, the neurons in the layer-1 are uncoupled. Otherwise,  we fix $\epsilon_{1} = 0.00128$ and $p=5$ (in nonlocal case).
%\subsection{Electrical coupling}
\par
\begin{figure*}[!ht]
		\centering
		\includegraphics[width=1.0\textwidth]{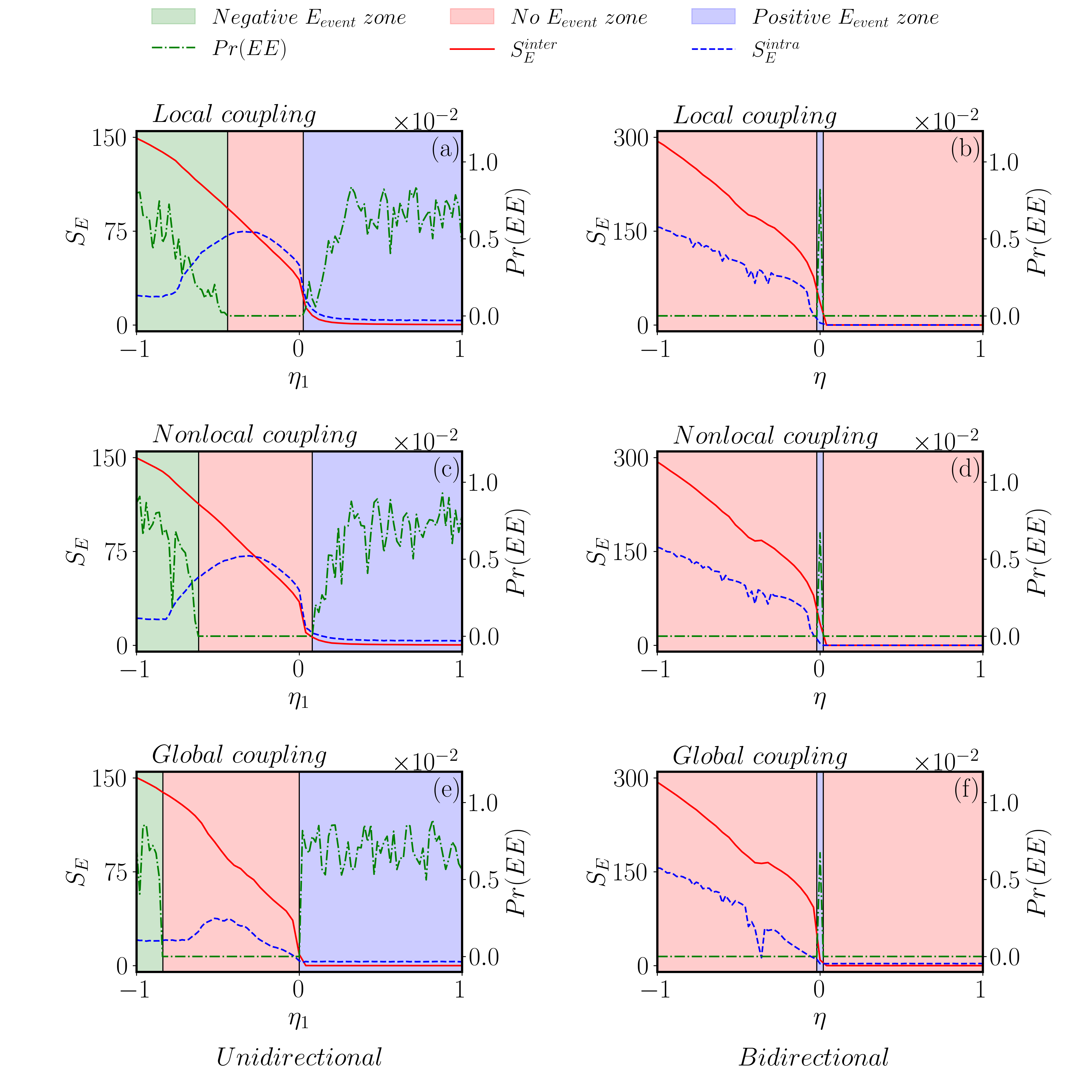}	
		\caption{$\eta_{2}$ $vs.$ $S_{E}$ and $\eta_{2}$ $vs.$ $Pr(EE)$ for unidirectional electrical interaction in first column (a,c,e) and for bidirectional electrical interaction in second column (b,d,f). First, second and third rows represent the analysis carried out for (a, b) local, (c, d) nonlocal and (e, f) global couplings in the intralayer networks of layer-1. For unidirectional and bidirectional configurations, we vary $\eta_{1}$ as in Fig.~\ref{se_ts_1} and $\eta_{1}=\eta_{2}=\eta$ as in Fig.~\ref{se_ts_3}. Here, we fix the range of nonlocal coupling at $p=5$. }
		\label{elc_conso}
	\end{figure*}

 $S^{inter}_{E}$, $S^{intra}_{E}$ and $Pr(EE)$ are plotted against $\eta$ for different coupling topologies of the intralayer and are shown in Fig.~\ref{elc_conso}. The columns (1 and 2) in Fig.~\ref{elc_conso} represent the analysis carried out for undirectional and bidirectional interactions respectively, while the rows (1-3) in Fig.~\ref{elc_conso} represent the analysis carried out for local, nonlocal and global couplings, respectively. In the unidirectional mode of interaction, despite different couplings adopted in layer-1, $S^{inter}_{E}$, $S^{intra}_{E}$ and $Pr(EE)$ follow the same patterns as in the uncoupled case. The only difference is the decrease in the range of $Pr(EE)$ in the negative spatial direction for nonlocal and global coupling. During bidirectional mode of interlayer interaction irrespective of different coupling schemes $S^{inter}_{E}$, $S^{intra}_{E}$ and $Pr(EE)$ remain the same as in the uncoupled case.
The results for chemical coupling are discussed in the Appendix. In both the cases, the dynamics is found to be robust and hence the emergence and mitigation of EE are also robust under different topologies in layer-1.

\par
\begin{figure*}[!ht]
	\centering
	\includegraphics[width=1.0\textwidth]{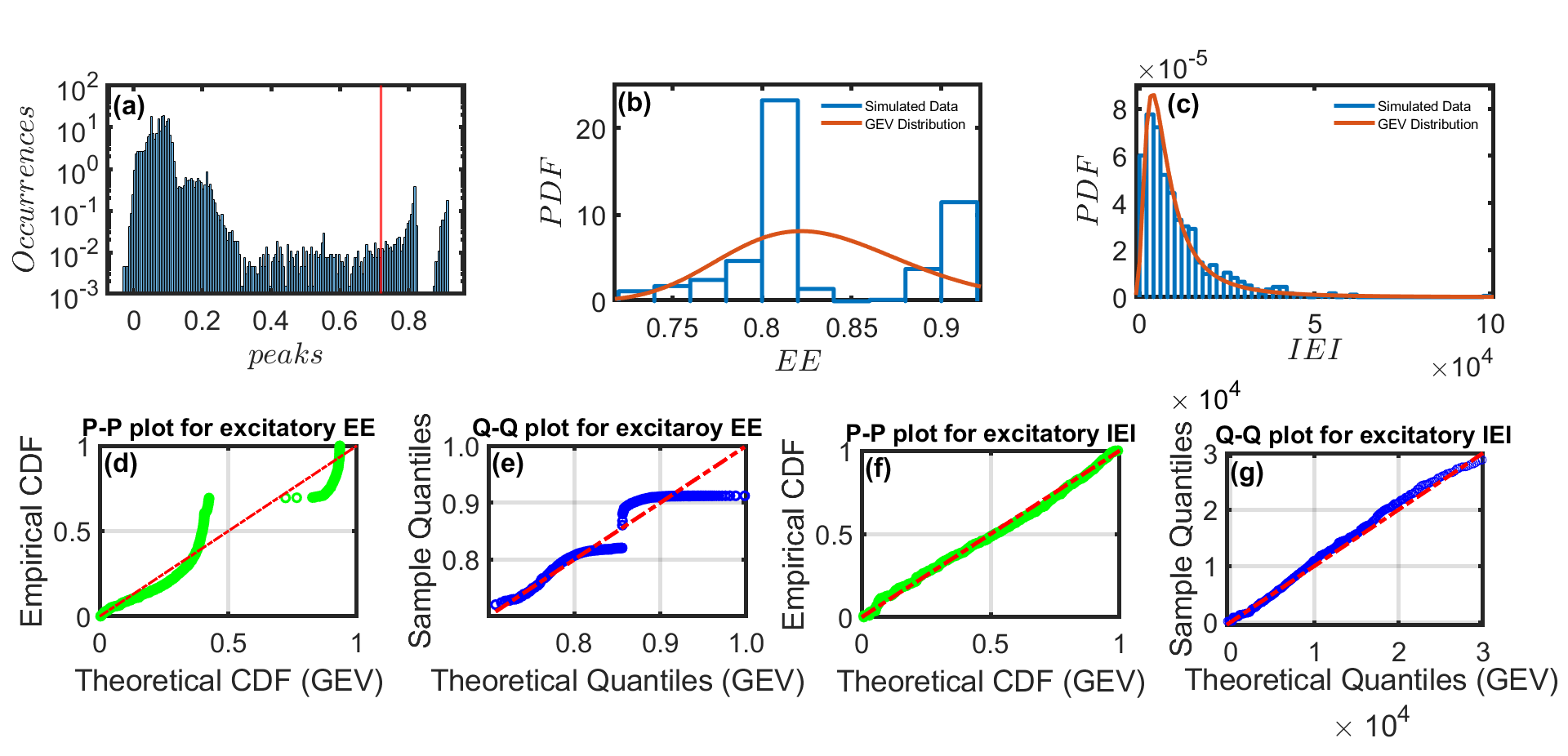}	
	\caption{\textbf{Statistics for electrical coupling (excitatory):} (a), (b) and (c) represents the histogram of peaks, the distribution fit of the EE and the distribution fit of the IEI. Plots ((d) \& (f)) and ((e) \& (g)) corresponds to distribution corrobation through Probability-Probability($P-P$) and Quantile-Quantile ($Q-Q$) calculations.}
	\label{el_exc_conso}
\end{figure*}

 We confirm the EE emerging in layer-1 through histograms, and appropriate statistical distributions. Specifically, for the electrical case, in Fig.~\ref{el_exc_conso}(a), we collect all the peaks and plot the histogram of peaks where we can find the peaks follow a long-tail behavior having occurrences beyond the threshold (red vertical line). In Fig.~\ref{el_exc_conso}(b), we statistically fit the EE data with the General Extreme Value (GEV) distribution, while in Fig.~\ref{el_exc_conso}(c), we statistically fit the GEV distribution to the interevent interval (IEI) data. Both EE and IEI get fitted with a non-Gaussian distribution confirming the nature of extremes. The GEV distribution is given by the mathematical form \cite{SNC}, 
\begin{eqnarray}
	G(x) &=& \dfrac{1}{\beta}\exp\Bigg(-\Big(1+\gamma\dfrac{x-\alpha}{\beta}\Big)^{-\frac{1}{\gamma}}\Bigg) \nonumber \\&& \times \Big(1+\gamma\dfrac{x-\alpha}{\beta}\Big)^{-\frac{1}{\gamma}-1}
\end{eqnarray}
for $\beta\neq0$ and $1+\gamma\dfrac{x-\alpha}{\beta} > 0$. Here, $\alpha>0$ is location parameter, $\beta >0$ is scale parameter, and $\gamma\neq0$ signifies the shape parameter. Depending on whether $\gamma$ is positive or negative, the distribution type varies as Fr\'echet (type II) and Weibull (type III) distributions, respectively. There is a third case scenario where the shape parameter $\gamma=0$ and corresponding distributions are called Gumbell (type I) and in this case, the distribution takes the following mathematical form,
\begin{eqnarray}
	G(x) = \dfrac{1}{\beta}\exp\Bigg(-\exp\Bigg(-\dfrac{x-\alpha}{\beta}\Bigg)-\dfrac{x-\alpha}{\beta}\Bigg).
	\label{dis2}
\end{eqnarray}
\par Further, we draw the probability-probability $(P-P)$ and quantile-quantile $(Q-Q)$ plots to corroborate the statistical fits. The  $P-P$ and $Q-Q$ plots tell how well the EE data and IEI data fit with GEV distributions. For $P-P$ and $Q-Q$ plots, we calculate cumulative distribution function (CDF) and Quantiles of EE, IEI and GEV data sets. The Figs.~\ref{el_exc_conso}(d) and \ref{el_exc_conso}(e) correspond to $P-P$ and $Q-Q$ plots of EE data and Figs.~\ref{el_exc_conso}(f) and \ref{el_exc_conso}(g) correspond to $P-P$ and $Q-Q$ plots of IEI data for excitatory coupling, respectively. From the obtained results, both the EE and IEI data fit with the GEV distribution.
\par We have also checked all these measures for the inhibitory case and the obtained results are similar to the excitatory case. We present the statistical measures for chemical coupling in Appendix.
\section{Mechanism}
\label{mechanism}
\par It is a well-known fact in the network theory that when the coupling strength is varied, the system macroscopically evolves from a disordered state to an ordered state.  In the present section, we study how EE emerge in layer-1 from a disordered state upon the variation of $\eta_{1}$. Since excessive synchronization is responsible for the emergence of epileptic seizures, we intend to observe how this excessive synchronization emerges in layer-1 from a disordered state while tuning $\eta_{1}$. This phenomenon is observed with the help of spatiotemporal, snapshots and frequency plots. For a single layer, in Ref. \cite{Ansmann} the authors have shown that proto-events are the responsible precursors for the emanation of EE. Here, proto-events denote the presence of more than 23 units in the excited state ($x_{i} > 0.6$). If such a proto-event occurs, subsequently all the 101 oscillators are synchronously excited at the same time producing EE. In the following subsections, we will discuss on how EE emerges and mitigates as we tune $\eta_1$ and $\eta_2$.
\subsection{FOR EMERGENCE}
\par
\begin{figure*}[!ht]
		\centering
		\includegraphics[width=1.0\textwidth]{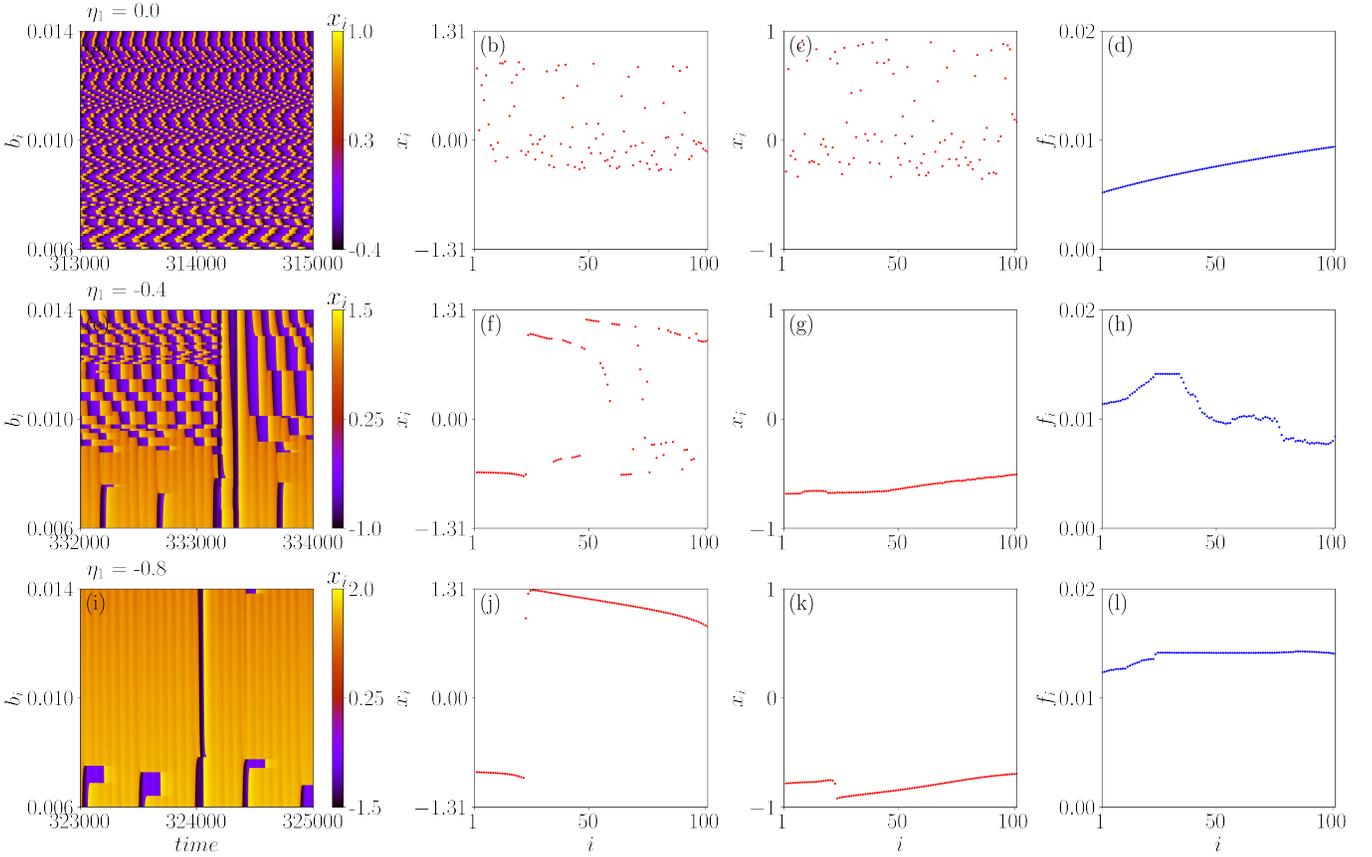}	
		\caption{The spatiotemporal, snapshots and frequency of layer-1 is displayed in column-1, columns-2,3 and column-4 while the rows denotes the plots for $\eta_{1} = 0.0, -0.4$ and $-0.8$ for unidirectional inhibitory coupling.}
		\label{sssf1}
	\end{figure*}

 For the electrical coupling in undirectional inhibitory case, in Fig.~\ref{sssf1}, we depict the spatiotemporal, snapshots and frequency plots of layer-1 in columns (1-4). The two snapshot columns (2 and 3) show the gradual emergence of proto-events and EE. Rows (1-3) in Fig.~\ref{sssf1} correspond to the plots for $\eta_{1} = 0.0, -0.4$ and $-0.8$, respectively. When $\eta_{1}=0.0$, layer-1 exhibits incoherent dynamics. The spatiotemporal plot in Fig.~\ref{sssf1}(a) indicates incoherent dynamics. From Figs.~\ref{sssf1}(b) and \ref{sssf1}(c) we can observe that the oscillators are oscillating in a random manner. When $\eta_{1}=-0.4$, the disordered state disappears and now the precursor for the emanation of EE also begins to be visible along with the sudden occasional complete synchronization. This can be viewed from Fig.~\ref{sssf1}(e). Between $t=333000$ and $t=333500$, we can observe the emergence of proto-events and subsequent excessive synchronization of oscillators. This can be further confirmed from Fig.~\ref{sssf1}(f) where we can observe that proto-events of 23 oscillators emerge (23 units crossing below -0.6) and Fig.~\ref{sssf1}(g) where all the oscillators are synchronized. Finally, increasing $\eta_{1}$ to $-0.8$, we can observe that the disordered state gets completely disappeared (Fig.~\ref{sssf1}(i)) and the EE emerge. %The change in inhibitory coupling $\eta_{1}$ leads to the excitation of the first 23 oscillators, which in turn stimulates the remaining units thereby exhibiting EE. 
From Fig.~\ref{sssf1}(j) we can observe that 23 units are beneath the value of -0.6 which corresponds to proto-events (similar to $x_{i} > 0.6$ as in Ref. \cite{Ansmann}) while in Fig.~\ref{sssf1}(k), all units are below $-0.6$ corresponding to EE.  For $\eta_{1}$ even up to 1.0, the same state persists. The frequency of oscillators is shown in fourth column of Fig.~\ref{sssf1} in which, the frequency of oscillators is in an increasing manner (Fig.~\ref{sssf1}(d)), is random (Fig.~\ref{sssf1}(h)) and evolves into two clusters (Fig.~\ref{sssf1}(l)). During EE, the frequencies of the oscillators split into two clusters, where the first 23 oscillators that are proto-events are in a hierarchical frequency while the remaining oscillators have the same frequency. This can be corroborated from the snapshots in Figs.~\ref{sssf1}(j-k).

\par In a nutshell, the unidirectional interlayer coupling drives the neurons in layer-1 to excite, as the signal from its counterpart in layer-2 arrives. Further, as the interlayer coupling strength gradually increases, the neurons in the layer-1 start to excite coherently. For higher values of the inhibitory coupling strength, when 23 or more neurons excite coherently, they act as the precursors (proto-events) for the synchronized excitation of all the neurons (emergence of EEs). We are observing the same phenomena for excitatory coupling as well. Further, for the unidirectional chemical coupling, the details about the emergence of extreme events are given in Appendix.
\subsection{FOR MITIGATION}
\par Layer-2 exhibits EE dynamics which is the source for the emergence of EE in layer-1. We discussed previously that the EE in layer-2 gets mitigated when the interaction is bidirectional. When the interlayer coupling is bidirectional, a constant excitation signal starts arriving in layer-2. On par with Ref. \cite{Shashang}, an excitation constantly arriving at layer-2 suppresses the EE. This is true even for a very small value of excitation (here small value of coupling strength). So, a very small value of $\eta$ directly affects layer-2 suppressing EE in the source layer-2.
\par
\begin{figure}[b]
		\centering
		\includegraphics[width=0.5\textwidth]{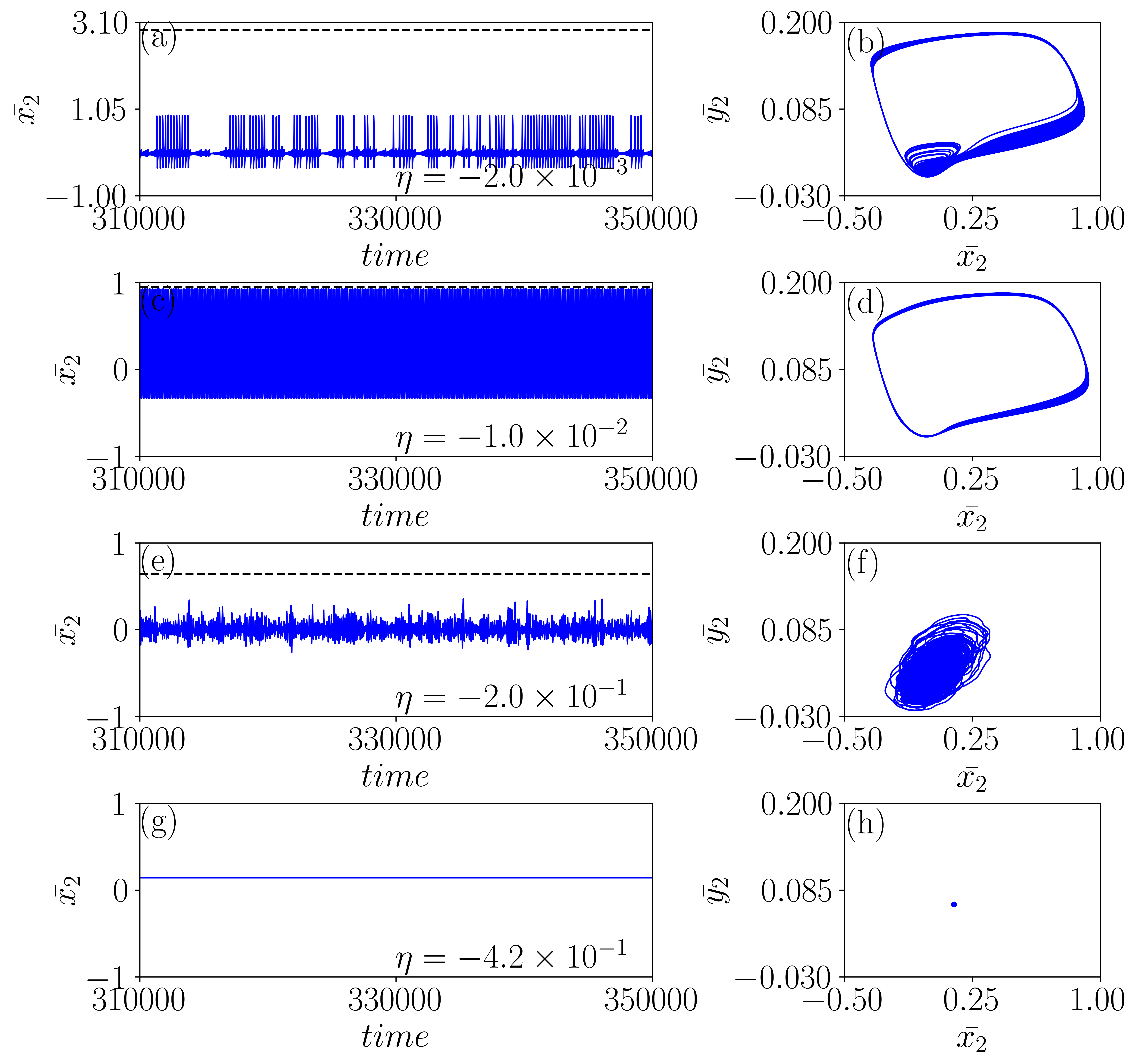}	
		\caption{The time series and phase portraits in column-1 and column-2 represent the suppression of the EE for inhibitory coupling in bidirectional electrical interaction. The rows represents the plots for the values of $\eta = -0.002$, $\eta = -0.01$, $\eta = -0.2$ and $\eta = -0.42$.}
		\label{m1}
	\end{figure}

 The gradual suppression of EE for electrical inhibitory coupling in bidirectional electrical interaction is shown in Fig.~\ref{m1}. Increasing $\eta$ to $-0.002$ increases the number of large amplitude peaks and there will be more number of long excursion trajectories from the bounded chaotic regime. The results can be viewed from Figs.~\ref{m1}(a) and ~\ref{m1}(b). Increasing $\eta$ to -0.01 further increases the large amplitude peaks (Fig.~\ref{m1}(c)) making it typical relaxation oscillation of FHN model (Fig.~\ref{m1}(d)). Further increasing $\eta$ to $-0.2$ transits into irregular oscillatory state (Figs.~\ref{m1}(e) and \ref{m1}(f)) and
then to steady state (Figs.~\ref{m1}(g) and \ref{m1}(h)) at $\eta$ to $-0.42$.
\newline
\par In essence, when the coupling is bidirectional, the layers receive the signals mutually. In this case, the dynamics in layer-2 are affected as a result of the incoming signal from layer-1, even before layer-1 is affected by layer-2. This is because, in layer-2, any signal affecting neurons can quickly spread to all the other neurons within the network due to its global coupling topology. This is true even for a very low value of the coupling strength. The dynamics get affected in such a way that the EE totally disappear. Since there are no EE in layer-2, there is no chance of EE propagation in layer-1. In a similar fashion, EE also gets mitigated for bidirectional excitatory electrical coupling. For bidirectional chemical coupling, the details about the route of mitigation of extreme events are provided in the Appendix.
\section{Conclusion}
\label{conclusion}
\par We have studied the emergence and mitigation of EE by multiplexing FHN neuronal network where two layers are connected through interlayer coupling. We have implemented interlayer electrical and chemical couplings for both unidirectional and bidirectional interactions. Differing from the results reported in \cite{Seager}, where the hubs and its handling capacity played a major role in the propagation and failure of EE, through this study, we report for the first time the dual role played by the interlayer coupling strength in both the emergence and mitigation of the EE depending on the direction of information transfer. Specifically, we have found that for unidirectional interaction, the interlayer coupling establishes intermittent synchronization leading to the emergence of EE in layer-1 from incoherent/disordered state, whereas in the case of bidirectional interaction, the transmission of information from the layer-1 to layer-2 acts as a constant DC voltage \cite{Shashang} which mitigates the EE in the layer-2 impacting the excitability of the system. We find that the propagation and mitigation of the EE occur irrespective of whether the neurons inside the layer or between the layers are synchronized or not. We have confirmed the emergence of the EE through necessary statistical measures and also studied the distribution nature of the extreme values. Interestingly, we found that our results are robust to different intralayer coupling topologies in layer-1. Emergence of EE occurs through the gradual disappearance of the disordered state interspersed with complete synchronization, while the mitigation happens through the de-excitation of oscillators from the excited state. 

\par In the literature, the propagation of FHN excitations and its failures have already been addressed in different monolayer network topologies, especially on how the degree contributes to the propagation failure of excitability in tree and random networks \cite{Kouvaris}, on how different spatiotemporal patterns arise as a result of addition of one long-range interaction in a regular ring network \cite{Thomas}, and on how global coupling strength affects the excitation propagation in a small-world network \cite{Isele}. But these studies were focused on the FHN excitation, propagation and failure only in monolayer networks. In the present work, the propagation of FHN synchronous excitation and its failure is studied between the layered networks. Further, in the only work on extreme events in multilayer networks in the literature, in Ref.~\cite{Seager}, the analysis is restricted to the transportation networks and the role of hubs, whereas in the present study, we have drawn our focus on the multiplex network with neuron models as nodes. So, the present \textbf{study} shed light on how epileptic seizures emerging in one layer propagate and affect the connected layers. Also, by uncovering the mechanism behind the emergence and mitigation of EE, we expect that our findings will be helpful in understanding the spread of occasional synchronization between layers in a broader manner. Our study has some scopes related to EEG study of brain activity \cite{Hramov, Stam}. Further, this direction of research may help us to understand the dynamical strategies for controlling the spread. The propagation of EE in a network of networks with different network topologies could be an interesting problem in near future. 
\section*{Acknowledgements}
	R.S. thanks Bharathidasan University, Tamil Nadu, India for providing, University Research Fellowship (URF). S.S. thanks the Science and Engineering Research Board (SERB), Department of Science and Technology (DST), Government of India for providing financial support in the form of National Post-Doctoral Fellowship (File No.~PDF/2022/001760).  The work of M.S. forms a part of a research project
	sponsored by the Council of Scientific and Industrial Research (CSIR) under Grant No. 03/1482/2023/EMR-II. D.G. is supported by DST-SERB Core Research Grant (Project No. CRG/2021/005894). Further, S.S. thanks Tapas Kumar Pal and Dhiman Das for fruitful discussions and valuable feedback on this manuscript, especially on the statistical part.
\section*{Appendix}
\subsection*{Statistics and Mechanism for chemical coupling}
\par In the main text, we have presented the statistical measures and the mechanism of propagation and mitigation of EE for the electrical interlayer coupling alone. Here we present the results when the interlayer coupling is chemical.
\par
\begin{figure*}[h]
	\centering
	\includegraphics[width=1.0\textwidth]{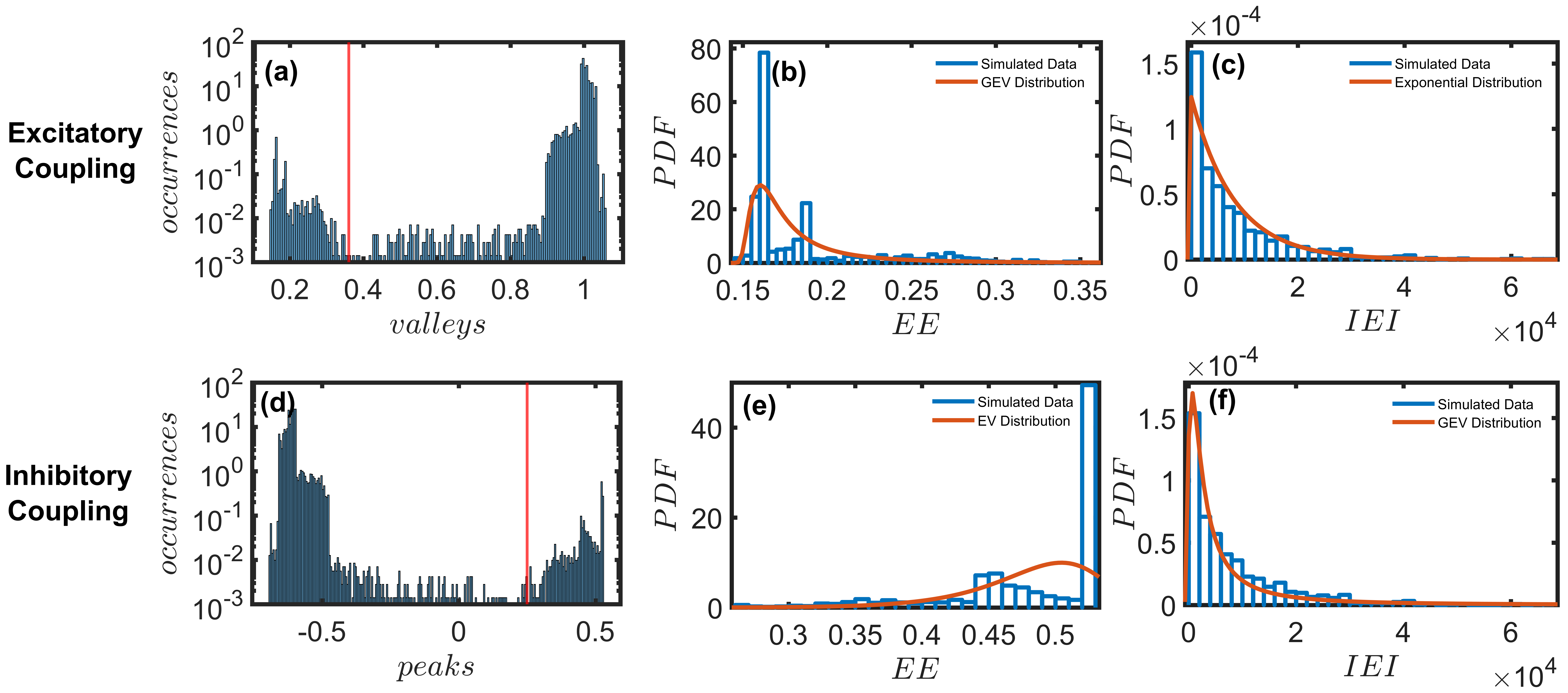}	
	\caption{\textbf{Statistics for chemical coupling:} (a),(d) represents the histogram of peaks, (b),(e) correspond to the distribution fit of the EE, and (c),(d) are the distribution fit of the IEI. Top row consists of the statistics for excitatory coupling while the bottom row is same for inhibitory coupling.}
	\label{ch_conso}
\end{figure*}

 We confirm the emergence of EE layer-1 through histograms, and appropriate statistical distributions for chemical coupling case. Similar to Fig.~\ref{el_exc_conso}, we plot in Fig.~\ref{ch_conso}, the histogram (Figs.~\ref{ch_conso}(a) $\&$ \ref{ch_conso}(d)), EE distribution (Figs.~\ref{ch_conso}(b) $\&$ \ref{ch_conso}(e)) and IEI distribution (Figs.~\ref{ch_conso}(c) $\&$ \ref{ch_conso}(f)).
\par The mathematical form of exponential distribution in Fig.~\ref{ch_conso}(c) is given by,
\begin{equation}
	\begin{array}{lcl} f(x,\mu)=
		\begin{cases} 
			\mu e^{-\mu x}; &  ~x \ge 0,  \\
			0; & ~x <0,
		\end{cases}
	\end{array}
\end{equation}
where $\mu$ is the rate parameter. The EV distribution in Fig.~\ref{ch_conso}(e) represents the Gumbell (type-I) distribution with $\gamma = 0$ given in Eq.~\ref{dis2}.

\begin{figure*}[h]
	\centering
	\includegraphics[width=1.0\textwidth]{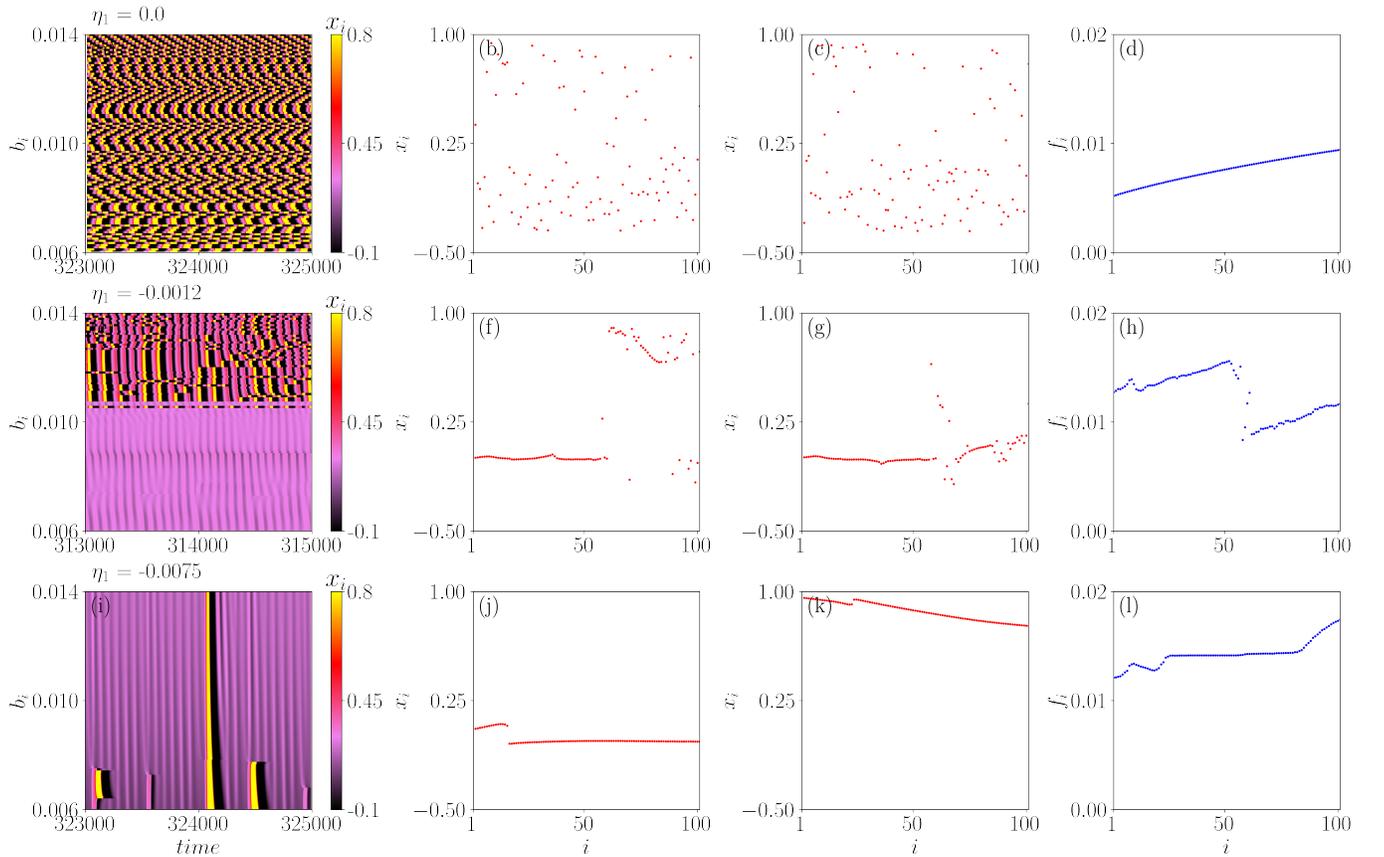}	
	\caption{The columns(1-4) represents the spatiotemporal, snapshots and frequency of layer-1 whereas the rows indicates the plots for $\eta_{1} = 0.0, -0.0012$ and $-0.0075$ for unidirectional inhibitory coupling.}
	\label{sssf3}
\end{figure*}	
\par To understand the mechanism for the case of chemical unidirectional inhibitory coupling, we present the spatiotemporal plots, snapshots and frequency of layer-1 in columns (1-4) of Fig.~\ref{sssf3}. The snapshots for non-extreme and EE are shown in columns-2 and 3. Rows indicate these plots for $\eta_{1} = 0.0, -0.0012$ and $-0.0075$. We can observe the incoherent dynamics for $\eta_{1} = 0.0$ from Fig.~\ref{sssf3}(a) with corresponding snapshots, Figs.~\ref{sssf3}(b) and \ref{sssf3}(c), showing disordered behavior of the oscillators. The coherent dynamics emerges along with the disappearance of incoherent dynamics (Fig.~\ref{sssf3}(e)) when $\eta_{1}$ is increased to $-0.0012$. Here more than half of the oscillators are synchronized and can be confirmed from the snapshots in Figs.~\ref{sssf3}(f) and \ref{sssf3}(g). Finally, EE emerges for $\eta_{1} = -0.0075$ where all the oscillators are occasionally in the excited state and also synchronized (Fig.~\ref{sssf3}(i)). In both the states (excited state and non-excited state) all the oscillators are synchronized (Figs.~\ref{sssf3}(j) and \ref{sssf3}(k)) but occasional synchronized excited state leads to EE in this case. Interestingly, in chemical coupling the EE occurs without the proto-events which means that even without precursor, all the oscillators occasionally gets excited synchronously at a particular time leading to the emergence of EE in layer-1. From the frequency plots in fourth column of Fig. ~\ref{sssf3} we can observe that the frequency of the oscillators is evolved into a three cluster state when the EE emerge in the system. For chemical inhibitory coupling, we observe the same phenomena.	
\par
\begin{figure}[h]
		\centering
		\includegraphics[width=0.5\textwidth]{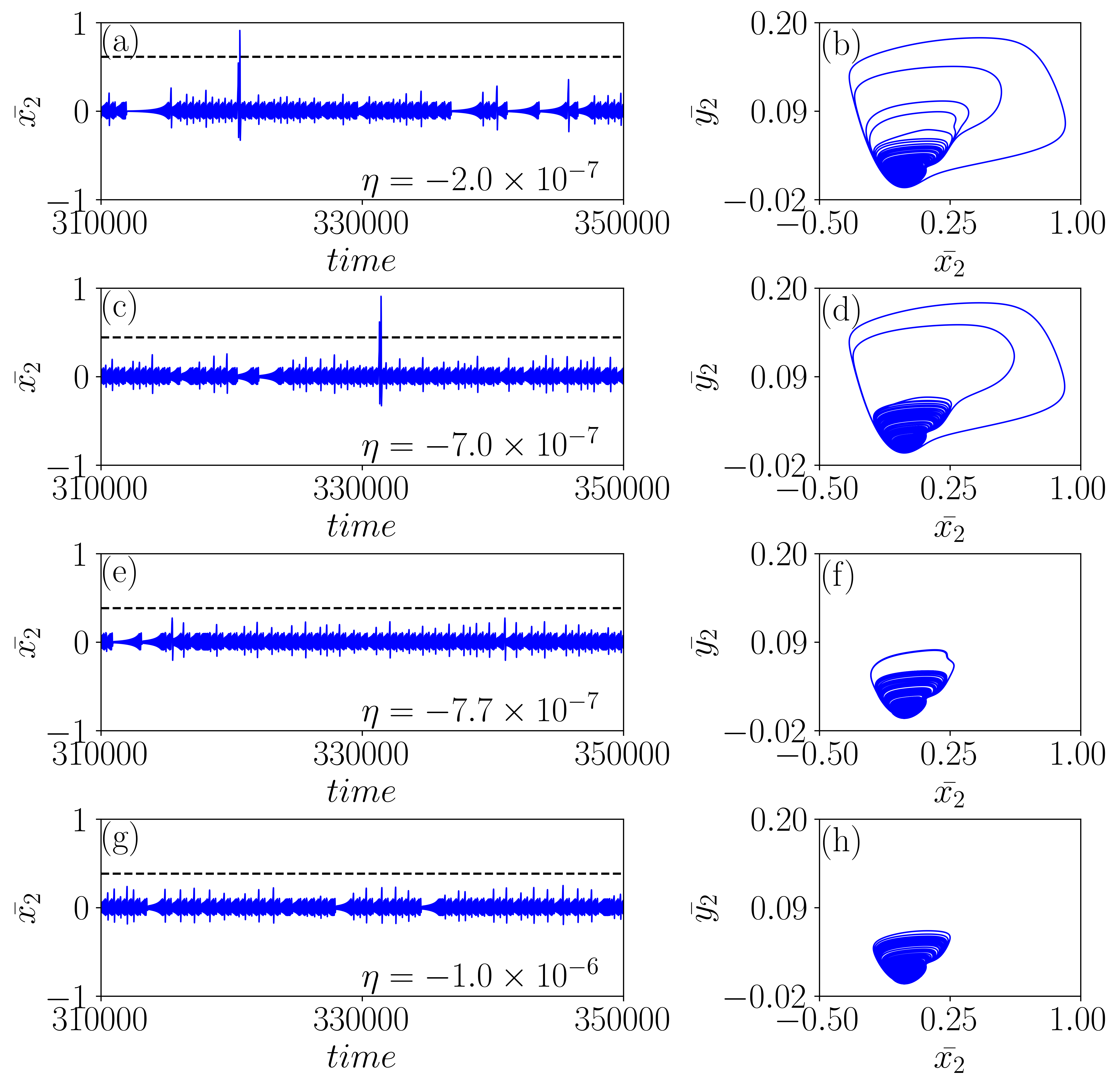}	
		\caption{The suppression of the EE for inhibitory coupling in bidirectional chemical interaction is shown via the time series and phase portraits in columns 1 and 2. The rows represents the plots for the values of $\eta = -2.0\times10^{-7}$, $\eta = -7.0\times10^{-7}$, $\eta = -7.7\times10^{-7}$ and $\eta = -1.0\times10^{-6}$.}
		\label{m3}
	\end{figure}

 The mitigation of the EE for inhibitory coupling in bidirectional chemical interaction is given in Fig.~\ref{m3}. Starting with $\eta = -2.0\times10^{-7}$, the EE occur which can be observed from the time series and phase portrait plots in Figs.~\ref{m3}(a) and ~\ref{m3}(b). The same dynamics occurs for $\eta=-7.0\times10^{-7}$ (Figs.~\ref{m3}(c-d)). Upon increasing $\eta=-7.7\times10^{-7}$, the EE get suppressed and only chaotic state prevails as there is no large amplitude peaks (in Fig.~\ref{m3}(e)) and long excursion trajectory (in Fig.~\ref{m3}(f)). Finally, increasing $\eta$ to $-1.0\times10^{-6}$, chaotic state appears (Figs.~\ref{m3}(g-h)) which persists up to $\eta=-1.0$. On the contrary, for excitatory coupling, extreme events get mitigated as in the electrical case. 
\par Results obtained for electrical coupling (in main text) and chemical coupling (in Appendix) are qualitatively the same. So it can be concluded that the propagation and mitigation of EE in the considered two-layer network depends only on the direction of information transfer.
\FloatBarrier

\end{document}